\documentclass[12pt,preprint]{aastex}

\bibpunct{(}{)}{;}{a}{}{,}

\def\gtapp
{\mathrel{\hbox{\raise0.3ex\hbox{$>$}\kern-0.8em\lower0.8ex\hbox{$\sim$}}}}
\def\ltapp
{\mathrel{\hbox{\raise0.3ex\hbox{$<$}\kern-0.75em\lower0.8ex\hbox{$\sim$}}}}
\newcommand{\hst}{{\sl HST\/}}

\slugcomment{Received ; Accepted }

\shorttitle{Environment of GRB hosts}
\shortauthors{Bornancini et al.}

\received{2003 December 16}
\begin{document}
\def\ale{\mathrel{\hbox{\rlap{\hbox{\lower4pt\hbox{$\sim$}}}\hbox{$<$}}}}
\def\age{\mathrel{\hbox{\rlap{\hbox{\lower4pt\hbox{$\sim$}}}\hbox{$>$}}}}

\title{The galaxy density environment of gamma-ray burst host galaxies$^{1,2}$}

\author{
Carlos~G.~Bornancini, H\'ector~J.~Mart\'{\i}nez\altaffilmark{3},
Diego~G.~Lambas\altaffilmark{3}}
\affil{
Grupo de Investigaciones en Astronom\'\i a Te\'orica y Experimental, IATE\\
Observatorio Astron\'omico, Universidad Nacional de C\'ordoba\\
Laprida 854, X5000BGR, C\'ordoba, Argentina}
\email{bornancini@oac.uncor.edu, julian@oac.uncor.edu, dgl@oac.uncor.edu}
\author{Emeric~Le Floc'h}
\affil{Steward Observatory, University of Arizona\\
933 Cherry Av., Tucson, AZ, 85721, USA}
\email{elefloch@as.arizona.edu}
\author{I.~F\'elix~Mirabel\altaffilmark{4}}
\affil{CEA/DSM/DAPNIA Service d'Astrophysique\\
F-91191 Gif-sur-Yvette, France}
\email{mirabel@discovery.saclay.cea.fr}
\and
\author{Dante~Minniti}
\affil{Department of Astronomy,
Pontificia Universidad Cat\'olica\\
 Vicu\~na Mackenna 4860, Casilla 306 Santiago 22, Chile}
\email{dante@astro.puc.cl}

\altaffiltext{1}{Based on observations with the Very Large Telescope, obtained at the European Southern Observatory in Chile under Proposal 67.B-0611(A)}
\altaffiltext{2}{\rm Based on observations made with the NASA/ESA {\sl Hubble Space Telescope (\hst)}, obtained at the Space Telescope Science Institute, which is operated by the Association of Universities for Research in Astronomy, Inc.\ under NASA contract NAS5-2655}

\altaffiltext{3}{
Consejo Nacional de Investigaciones Cient\'\i ficas y T\'ecnicas (CONICET),
Avenida Rivadavia 1917, C1033AAJ, Buenos Aires, Argentina.}
\altaffiltext{4}{Instituto de Astronom\'{\i}a y F\'{\i}sica del Espacio, cc67, suc28. 1428 Buenos Aires, Argentina.}

\begin{abstract}

We analyze cross-correlation functions between Gamma-Ray Burst (GRB)
hosts and surrounding galaxies. We have used data obtained with the Very Large
Telescope at Cerro Paranal (Chile), as well as public Hubble Space Telescope data.
Our results indicate that Gamma-Ray Burst host galaxies do not reside in high galaxy
density environments. Moreover, the host-galaxy cross-correlations show a relatively low amplitude. Our results are in agreement with the cross-correlation function between star-forming galaxies and surrounding objects in the HDF-N.

\end{abstract}



\section{Introduction}
The origin of cosmological Gamma-Ray Bursts (GRBs) is still one of the
outstanding problems in modern astronomy \citep{vanpa97}.
Over the past half decade, the discovery of localized transients in the small error boxes of GRBs has led to intense multiwavelength campaigns and many advances have been made in understanding the nature of the bursts and their afterglows throughout the electromagnetic spectrum (see for instance \citet{vanpa00}).\\
The scenario for the origin of GRBs include either the merging of two collapsed objects such as black holes or neutron stars merge \citep{nara92} or the cataclysmic destruction of massive stars (supernovae or hypernovae) \citep{woos93,macwoos99}.
In the first scenario, GRBs can occur long after the star formation episodes in the nucleus since the merging characteristic time scale is as long as $\gtapp$ 1 Gyr. In the second scenario the host galaxies are likely to be on-going significant star formation.

\citet{bloom02} analyzed the observed offset distribution of Gamma-Ray Burst from their host galaxies, and found a strong connection of GRB localization with the UV light of their host galaxies. This provides a significant observational evidence for the correlations between GRBs and star-forming regions.

The characteristics of the physical environment surrounding the bursts may provide strong constrains for the origin of these events.
As suggested by \citet{macwoos99} in the collapsar model, GRBs could be
produced by rotating massive stars
in which the accretion of a helium core leads to the prompt formation of a
black hole. Low metallicity in the stellar envelope reduces the mass loss and
inhibits the loss of angular momentum by the star.
Star formation in regions of low metallicity are likely to generate GRBs so that dwarf and sub-luminous galaxies would be preferred host of these events (see for instance \citet{floch03} ).

This model is supported by the Ly$\alpha$ emission from GRB host galaxies which show a preference to be metal poor and with on-going star formation in a dust poor environment \citep{Fynbo02,Fynbo03}.

Analyzing the properties of the host galaxies of GRBs, \citet{floch03} found that GRB host galaxies are sub-luminous in the $K$-band and exhibit very blue colors, not comparable to the luminous star-burst and/or reddened sources observed at high redshifts in the infrared and sub-millimeter deep surveys.

There is new evidence that at least some Gamma-Ray Bursts are related to supernovae: the association of GRB 980425 with the peculiar Type Ic  SN 1998bw \citep{gala98}, and also the recent spectroscopic discovery of the
SN 2003dh associated with GRB 030329 \citep{hjorthnat,stanek03}.
These observations provide a very solid confirmation of the association of some GRBs with the death of massive stars in strong star formation regions.

At the present it is unknown if GRB host galaxies are preferentially located in dense environments, or if there is any correlation between the local density of galaxies and the presence of a GRB. So far, there are only two works on the environment of GRB host galaxies.
\citet{Fynbo02} studying the Ly$\alpha$ emission of two GRB fields (GRB 000301C \& GRB 000926),
 found a number of galaxies at the same redshift of GRB hosts without signs of overdensity at small scales. Moreover, the lack of blank fields at a similar depth prevented these authors to conclude if GRB hosts reside in overdense regions.
From an analysis of photometric redshifts of galaxies in the field of GRB 000210, \citet{goro03} found that there is no obvious concentration of galaxies around the host. Another case, the afterglow of GRB 980613 was located close to a very
compact object inside a complex region consisting of star-forming knots
and/or interacting galaxy fragments \citep{Hjorth02,djo03}. If these host galaxies are associated with the underlying large scale structure of the universe, then they should show similar galaxy clustering properties of normal galaxies.\\
In this paper we examine the nature of the density enhancement
of galaxies around GRB host galaxies by studying the cross-correlation function between GRB hosts and the surrounding galaxy distribution.

This paper is organized as follows: Section 2 describes the sample of GRB host galaxies analyzed, the source extraction and photometry techniques. We
analyze galaxy source counts in Section 3. Section 4 deals with the angular-cross correlation analysis of galaxies around GRB host galaxies. Finally we discuss our results in Section 5.

\section{Data}

The near-IR data were obtained with the Infrared Spectrometer And Array
Camera ({\tt ISAAC}) on the Very Large Telescope (VLT) at Paranal, (Chile)
between March 2000 and September 2001 under photometric conditions. A $K_{s}$ filter (2.0-2.3 $\micron$) was used. The focal lens configurations resulted in a respective pixel size of 0\farcs148.
All the images were obtained under optimum seeing conditions with FWHM in the range 0\farcs6 and 1\farcs5.
For the {\tt ISAAC} observations, we reached a total on-source integration
time of 1\,hour per object.
A full description of sample selection, observations and image reductions is given in \cite{floch03}.
Our sample sources is listed in Table~\ref{tab1}, together with the position, spectroscopic redshift determinations taken from the literature, the $K_{s}-$band magnitudes of the host galaxies and the 1.5$\sigma$ limiting magnitude per field.\\
The optical GRBs sample used in this work consist in 19 \hst/STIS GRB host galaxies images.
These observations were taken from Hubble Space Telescope (\hst) STIS
imaging data from the Cycle 9 program GO-8640 ``A Survey of the Host
Galaxies of Gamma-Ray Bursts'' (\citep{holland698} \footnote{Data and further
information available at
\texttt{http://www.ifa.au.dk/\~{}hst/grb\_hosts/index.html}}).
Images were obtained with the {\tt 50CCD} filter (clear, pivot
$\lambda _{o}$=5835 \AA, \ hereafter {\tt CL}) and {\tt F28X50LP}
(long pass, pivot $\lambda _{o}$=7208 \AA, \ hereafter {\tt LP}).
The data was pre-processed using the standard STIS pipeline and
combined using the DITHER (v2.0) software \citep{Fruchter2002.1} as
implemented in IRAF\footnote{Image Reduction and Analysis Facility
(IRAF), a software system distributed by the National Optical
Astronomy Observatories (NOAO).}  (v2.11.3) and STSDAS (v2.3). The
STIS images were drizzled using `pixfrac=0.6' and `scale=0.5' (giving
a final pixel size of 0\farcs0254).
Images were selected for having an optical and/or radio bright afterglow with the deepest observations available in the {\tt 50CCD} images.
For the STIS zero-points, we adopted the values found by \citet{gard00}
for the  HDF$-$South in the  AB system \citep{oke71}. The zero-points used were therefore ZP$_{CL}=26.387$ and ZP$_{LP}=25.291$.
We have measured a PSF from the STIS images with a FWHM$=0.083-0.085$, similar to that obtained by \citet{gard00} in the  HDF$-$S STIS analysis. We adopt this value for images without stars present in the field.
\hst/STIS sources are listed in Table~\ref{tab2}, for the {\tt 50CCD} filter, together with the position, spectroscopic redshift determinations taken from the literature, the {\tt CL} band magnitudes of the host galaxies, the total on-source exposure time and the 2$\sigma$ limiting magnitude per field.\\

\subsection{Source extraction and photometry}

For object detection and photometry we used the SExtractor software package version $2.1$ \citep{bertin96}.
For the {\tt ISAAC} images, the source extraction parameters were set such that, to be detected, an object must have a flux in excess of $1.5$ times the local background noise level over at least $N$ connected pixels, where $N$ was varied according to the seeing conditions (about $\sim$10$-$15 connected pixels).

SExtractor's {\tt MAG\_BEST} estimator was used to determine the magnitudes
of the sources; this yields an estimate for the total magnitude
using first Kron's (1980) moment algorithm.

In this work we choose all objects (galaxies) with stellaricity
index $<0.8$, for the {\tt ISAAC} images.
The result of the detection process was inspected visually in order to ensure that no obvious objects were missed, and that no false detections were entered into the catalogs. Saturated objects and objects lying in the image boundaries were rejected from the catalogs. The final effective field of view is $2\farcm2\times2\farcm2$, after spurious detections near boundary regions are rejected.

For the \hst/STIS images, after extensive tests with SExtractor using different input parameters, it became clear that no single set of parameters could satisfy to create a catalog that would reach the faintest depth for isolated objects and faint companions of bright galaxies.
This is a standard problem with SExtractor and at the moment an ideal solution does not exists.
For example, \citet{caserta00} and \citet{gard00} ran SExtractor two times with different detection thresholds. In this way, however, it is not easy to achieve consistent estimates of the fluxes with different thresholds. For these reasons, we have used a detection threshold corresponding to an isophote of $2\sigma$ and objects were required a minimum area of 10 connected pixels above this threshold and a stellaricity index ${\tt CLASS\_STAR}<0.95$, for an object to be a galaxy. After this, spurious objects in correspondence with stellar diffraction spikes and spurious galaxies detected from the spiral arm fragmentation of nearby galaxies were removed from the final catalog. The final effective field of view for the STIS images is $25\farcs6\times25\farcs6$.

\section{Near-IR and Optical Galaxy Number Counts}

\begin{figure}
\plottwo{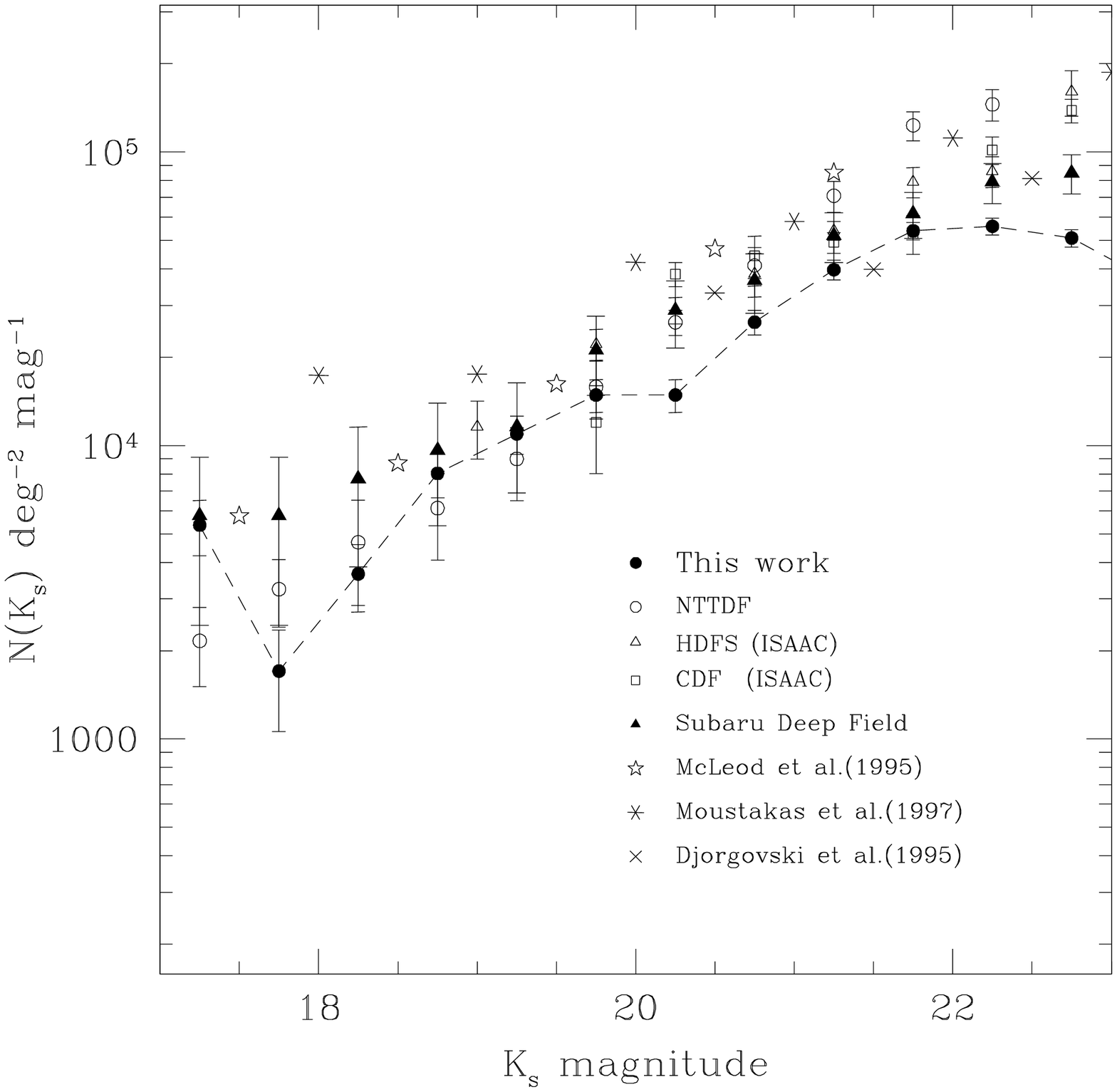}{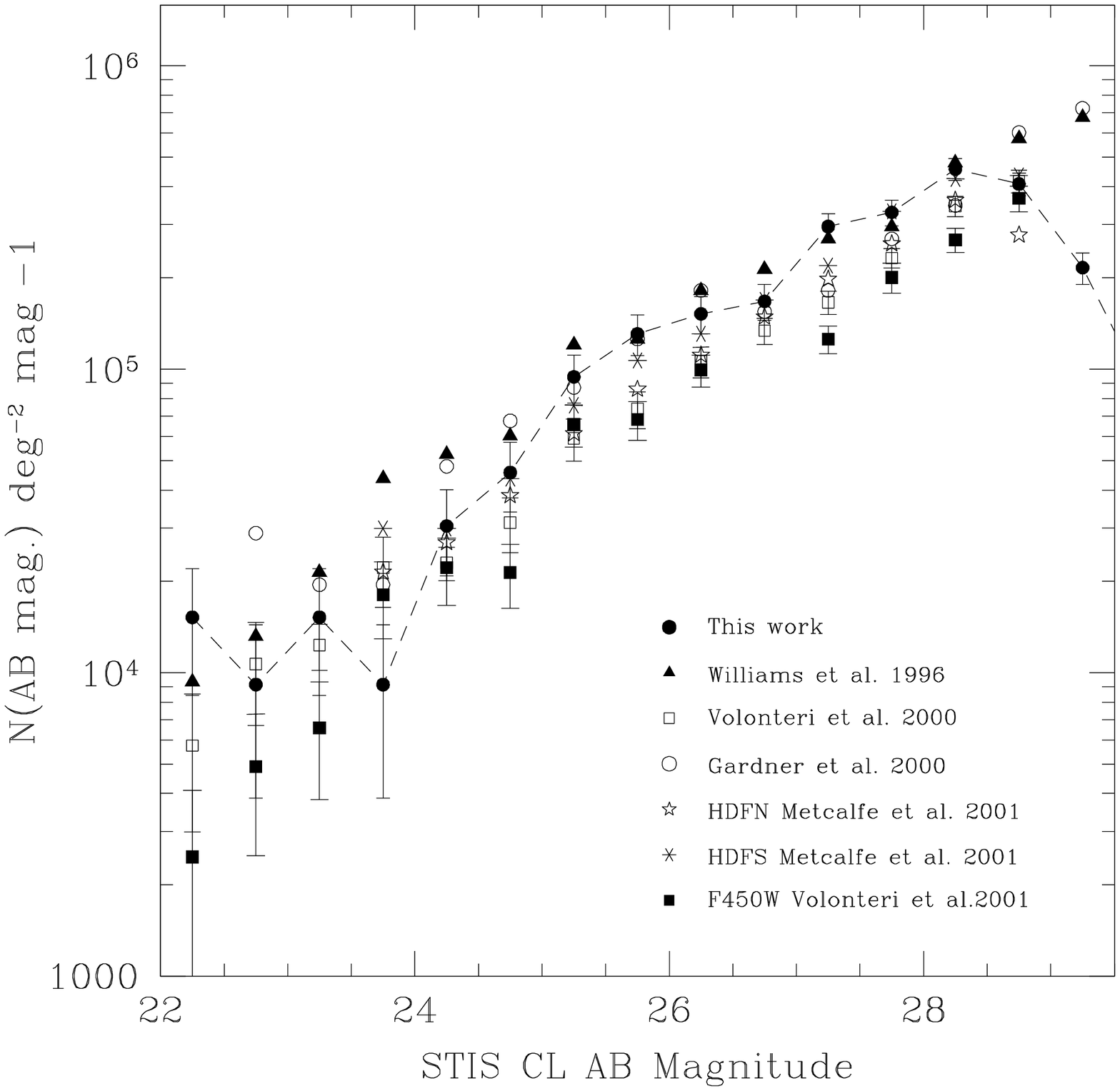}
\caption{Near-IR number counts in the {\tt ISAAC} images (left panel) and optical number counts in \hst/STIS images (right panel) for GRB fields and
determinations from the literature. }
\label{fig1}
\end{figure}

In order to check for completeness in the sample analyzed, we have computed the mean number of galaxies per unit area for the fields.
The results are plotted in Figure~\ref{fig1}. Error bars were estimated using Poisson counting statistics on the raw galaxy counts. In Figure~\ref{fig1} (right panel), we compare our determinations with the $K_{s}$
number counts in the Chandra Deep Field (CDF) and in the Hubble Deep Field South (HDF$-$S) \citep{sara01} and with determinations from the NTT Deep Field \citep{sara99}.
We also compare our counts to those of other $K$-band surveys from \citet{mousta97}, \citet{mcleod95}, \citet{djo95}, as well as from the Subaru Deep Field \citep{tota01}.
Our $K_{s}$ number counts derived here are systematically lower than those from the literature, in the range $20<K<22$, which could be an indication of lack of
 many faint galaxies in these GRBs fields.
We have also computed the differential galaxy counts in the {\tt 50CCD} ({\tt CL}) images, (see Figure~\ref{fig1}, left panel). Error bars were estimated using Poissonian errors. We compare our determinations with the {\tt 50CCD} number counts from  HDF-S STIS imaging \citep{gard00}. For comparison, we plot the  WFPC2 HDF-N galaxy counts in $F606W$ from \citet{willi96} from \citet{volo00}, from the HDF$-$S and HDF$-$N and in F450W from \citet{met01}. These results are fully consistent with those in the literature.

\section{Angular cross-correlation analysis}

In this section we analyze the clustering of galaxies around GRB host galaxies.
We compute the angular two point cross-correlation function $\omega(\theta)$
between the GRB host galaxies and the galaxies in their fields for the {\tt ISAAC} near-IR images and for the \hst/STIS images (tracer galaxies).
We have used the following estimator of the angular cross-correlation function, \citep{peebles80}:

\begin{equation}
\omega(\theta)=\frac{n_R}{n_G}\frac{DD(\theta)}{DR(\theta)}-1
\end{equation}
where $n_G$ and $n_R$ are the numbers of galaxies in the sample and in a random sample respectively, $DD(\theta)$ is the number of real pairs host-galaxy
separated by an angular distance in the range $\theta, \theta+\delta \theta$,
and $DR(\theta)$ are the corresponding pairs when considering the random
galaxy sample. 
We have also computed the auto-correlation function of the tracer galaxies in 
all fields analyzed which serves to compare the relative clustering strength 
around GRB hosts, and around typical galaxies.

We estimate correlation function error bars using the field-to-field
variation:
\begin{equation}
\sigma^2 =
\frac{1}{N-1}\sum_{i=1}^{N}\frac{{DR_i(\theta)}}{{DR(\theta)}}\left[\omega_i(\theta)-\omega(\theta)\right]^2
\end{equation}
where the subscript $i$ stands for each 
individual frame.
The field-to-field errors are $1\sigma$ standard
deviations of the correlation function between fields,
and are inverse variance-weighted to account for the different numbers
of sources on each field.
\citet{Myers} conclude that these variations provide suitable 
estimates of correlation uncertainties  accounting for
cosmic variance and differences of image photometric zero-point calibrations.

The finite area of the images implies a systematic 
amplitude offset known as the integral 
constraint {\tt $C$}, such that $\omega_{obs}(\theta)=A(\theta^{1-\gamma}-C)$.
The integral constraint can be computed numerically using a random-random sample:

\begin{equation}
C=\frac{\sum RR(\theta)\theta^{1-\gamma}}{\sum RR(\theta)},
\end{equation}

where $RR(\theta)$ is the number of random pairs of objects with
angular distances between $\theta$ and $\theta+\delta\theta$.
For our sample geometry, and assuming $\gamma=1.8$, we find 
$C=0.062~{\rm arcsec}^{-0.8}$, for {\tt ISAAC} images and $C=0.13~{\rm arcsec}^{-0.8}$ for \hst/STIS images.\\

\begin{figure}
\plottwo{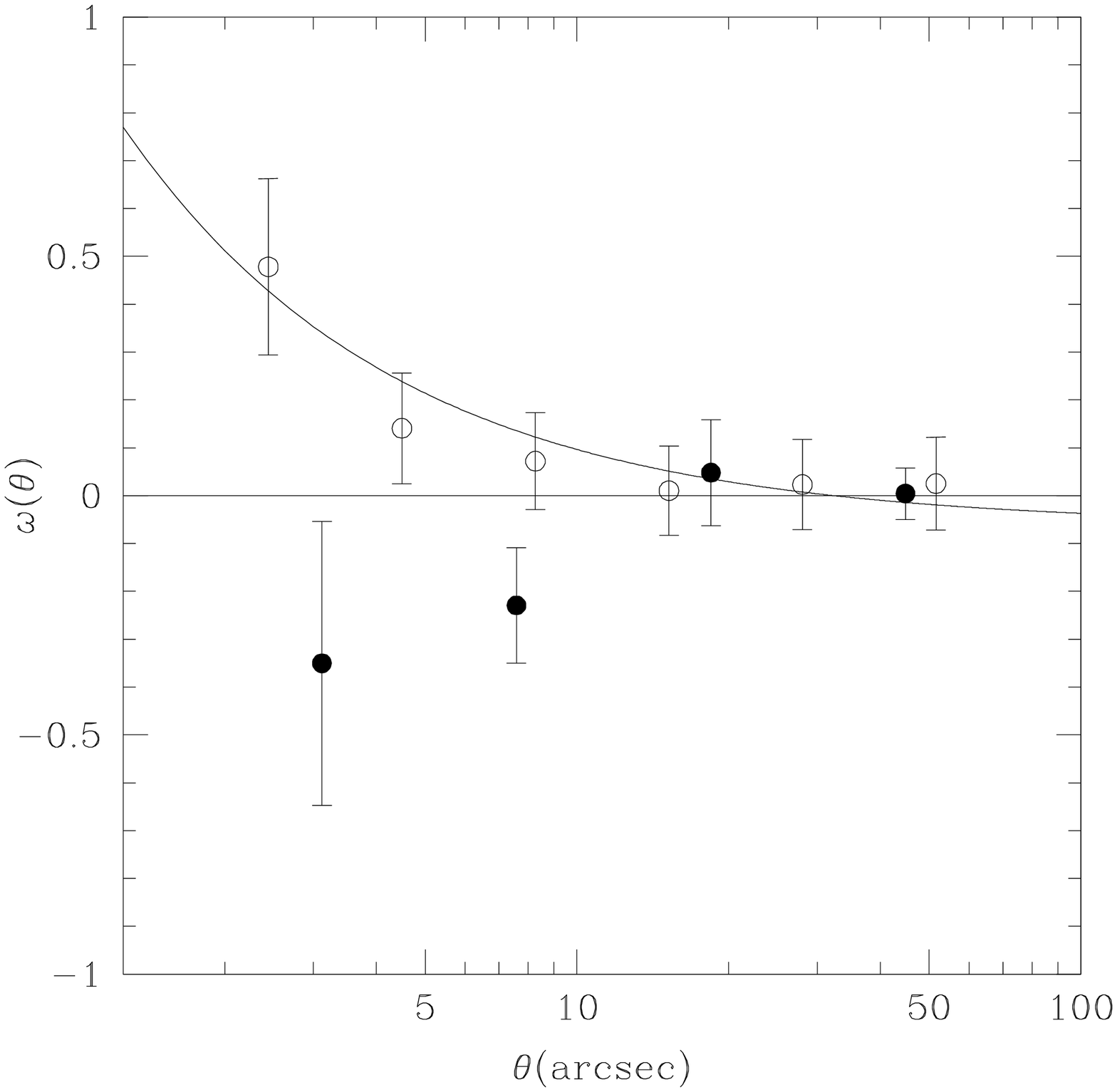}{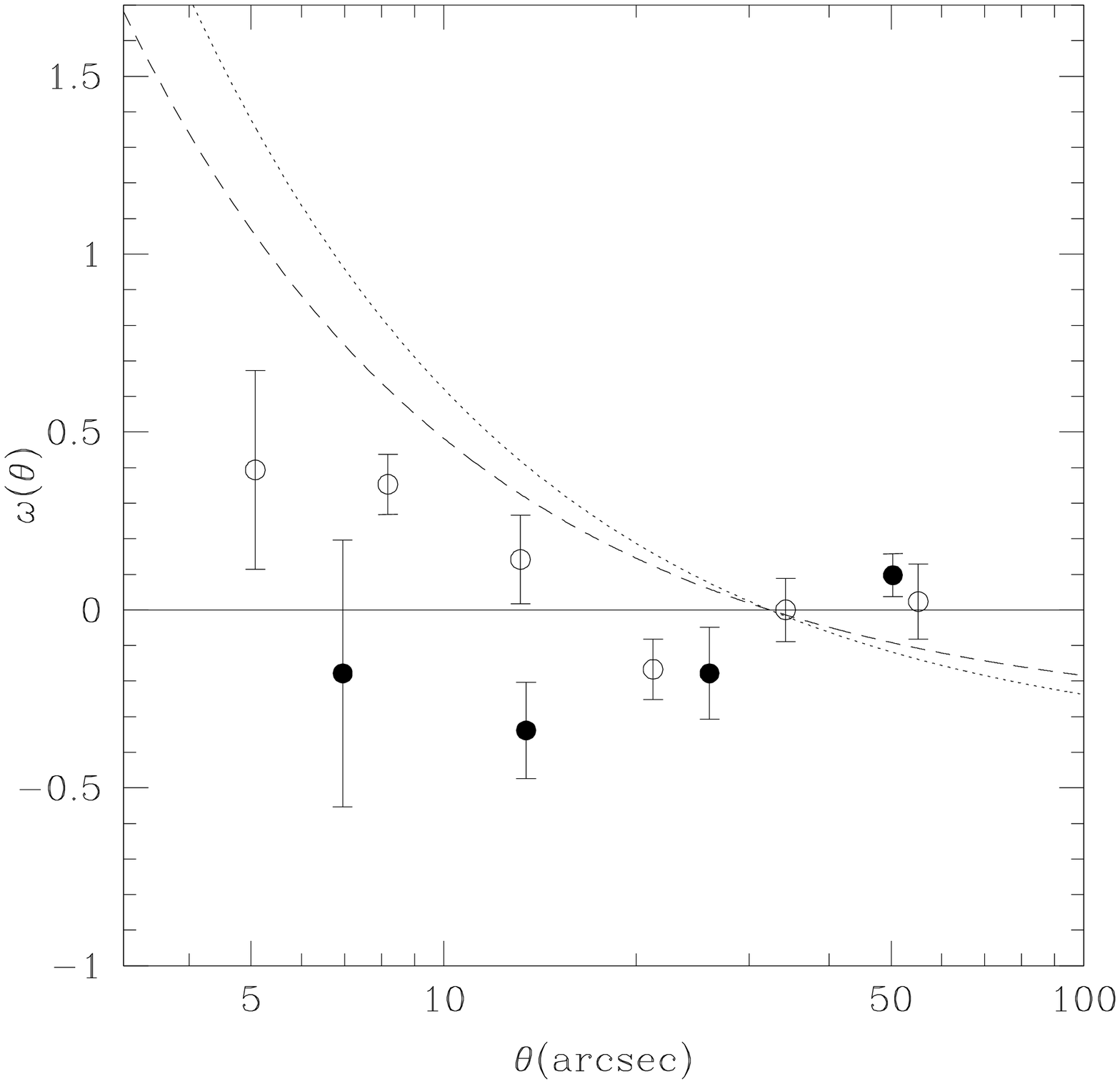}
\caption{Two point cross-correlation functions GRB-tracer galaxies (filled circles) and auto-correlation functions of tracers galaxies (open circles). Left panel: 
K$_{s}<21.5$. Solid line is the best power law fit to auto-correlation for
tracer galaxies with $K_{s}<21.5$. Right panel:  K$_{s}<19$. 
The dotted line correspond to K$<19$ radio galaxy-galaxy correlation function 
taken from \citet{best}. 
The dashed line corresponds to
USS-galaxy correlation function 
for tracers with $18<K<19$ 
taken from \citet{bornan04}}
\label{VLT1}
\end{figure}

\begin{figure}[tbp]
\epsscale{0.5}
\plotone{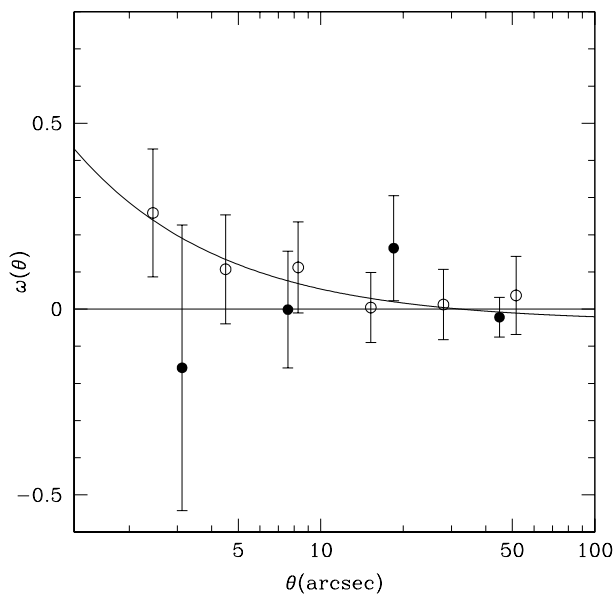}
\caption{Same as previous figure but tracer galaxies restricted to 19.5 $<K_{s}<21.5$ }
\label{VLT2}
\end{figure}

In Figure~\ref{VLT1} and ~\ref{VLT2} we show the resulting host-galaxy two-point cross-correlation
functions for three samples of tracer galaxies: $K_{s}<21.5$, $K_{s}<19$, and  $19.5<K_{s}<21.5$ in the {\tt ISAAC} images. These magnitudes limits are consistently with the K$-$ band photometry and spectroscopic redshifts determinations of field galaxies (see Figure 2 in \citet{floch03}).
In the same figures we have computed the auto-correlations functions of the tracer galaxies in the same magnitudes range.
The amplitude of the auto-cross correlation function is $A=(0.8\pm0.3)~{\rm arcsec}^{0.8}$, for the all tracer galaxies sample with $K_{s}<21.5$; and 
$A=(0.56\pm0.33)~{\rm arcsec}^{0.8}$ for tracers galaxies with $19.5<K_{s}<21.5$.

\begin{figure}[tbp]
\plotone{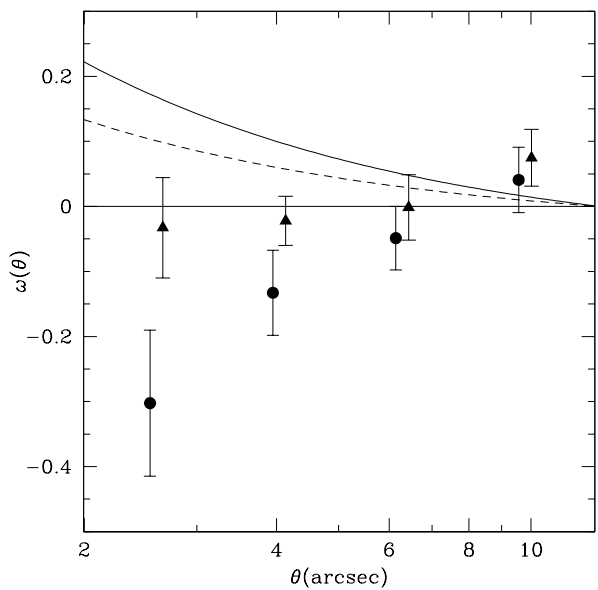}
\caption{GRB-tracer galaxies (${\tt CL}<28$) cross-correlation function in the \hst/STIS images and for all GRB hosts quoted in Table~\ref{tab2} (filled triangles), the filled circles correspond to GRB hosts in the redshift
range $0.7<z<2$. 
Solid line is the best power law fit to auto-correlation function for
tracer galaxies with ${\tt CL}<28$. Dotted line corresponds to auto-correlation function for GRB hosts in the redshift range $0.7<z<2$  }
\label{fig5}
\end{figure}

We have also performed this computation for the \hst/STIS images which are shown in Figure \ref{fig5}.
In this figure we can appreciate the host-galaxy cross-correlation function in the {\tt 50CCD} images for tracer galaxies with ${\tt CL}<28$ around all host galaxies quoted in Table~\ref{tab2} (filled triangles).\\

We have also explored different intervals of magnitudes and the effect on the results of excluding hosts that are either too distant ($z>2$), or relatively close ($z<0.7$) in order to avoid very faint galaxy images and large angular scales.
In Figure \ref{fig5} we show the resulting host-galaxy cross-correlation
functions for tracer galaxies with ${\tt CL}<28$ around host galaxies with spectroscopic redshift determinations in the range $0.7<z<2$ (Filled circles).
The amplitude of the auto-cross correlation function is $A=(0.5\pm0.2)~{\rm arcsec}^{0.8}$, for the all tracer galaxies sample with ${\tt CL}<28$; and 
$A=(0.30\pm0.17)~{\rm arcsec}^{0.8}$ for tracers galaxies around GRB host galaxies with spectroscopic redshifts in the range $0.7<z<2$ in the same magnitude range.

In this figure we have not included GRB 990705, because this GRB host is located behind the outskirts of the Large Magellanic Cloud (LCM) and it is expected a stronger contamination in the SExtractor point source identification.

We can see here that there is not a significant 
GRB host-galaxy cross-correlation amplitude.
In similar analyzes, \citep{best, bornan04}
, found that radio galaxies and 
Ultra Steep Spectrum radio sources with comparable redshifts than 
our GRB sample ($z \simeq 1$). 
show  significant cross-correlations with neighboring galaxies
(shown as dotted and dashed lines in Figure~\ref{VLT1}, right panel)
indicating that these sources are
likely to reside in proto-cluster environments. By contrast, 
the lack of a cross-correlation signal in our GRB fields provides 
clear evidence 
that the neighborhood of GRB hosts is of significant lower galaxy
overdensity than groups and clusters. 
This is reinforced by the fact that GRB-galaxy cross-correlations have a lower amplitude than the galaxy auto-correlation function in these fields 
for the two limiting magnitudes analyzed. 

We have restricted to tracer galaxies beyond an angular radius $\theta_{p}>2\farcs0$ from the host. We adopt this value to avoid the numerous knots or galaxy fragments, observed in some host complex systems and the surrounding galaxies.
It can be appreciated in these figures a significant anti-correlation signal
 between GRB hosts and surrounding galaxies, indicating that GRB hosts reside in regions strongly biased to low local galaxy densities.\\

\begin{figure}[tbp]
\epsscale{0.5}
\plotone{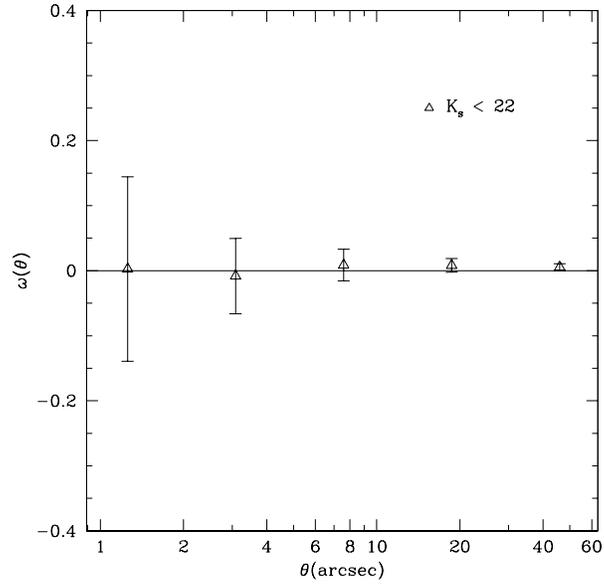}
\caption{Star auto-correlation function test}
\label{starVLT}
\end{figure}

We have also tested for possible bias in the detection
of objects near the edges of the frames of GRBs images.
We have computed auto-correlation analysis for objects with ${\tt CLASS\_STAR}>0.8$ for the {\tt ISSAC} images and those with ${\tt CLASS\_STAR}>0.95$ for the \hst/STIS images,
identified in our images as point sources (stars).
We find that stars are uncorrelated on the sky, as shown in Figure \ref{starVLT} for $K_{s}<20$. We find similar values for ${\tt CL}<28$ and ${\tt CL}<26$ in the \hst/STIS images. The lack of signal for these
samples indicates the absence of significant systematic effects in our analysis.\\

\begin{figure}[tbp]
\epsscale{0.5}
\plotone{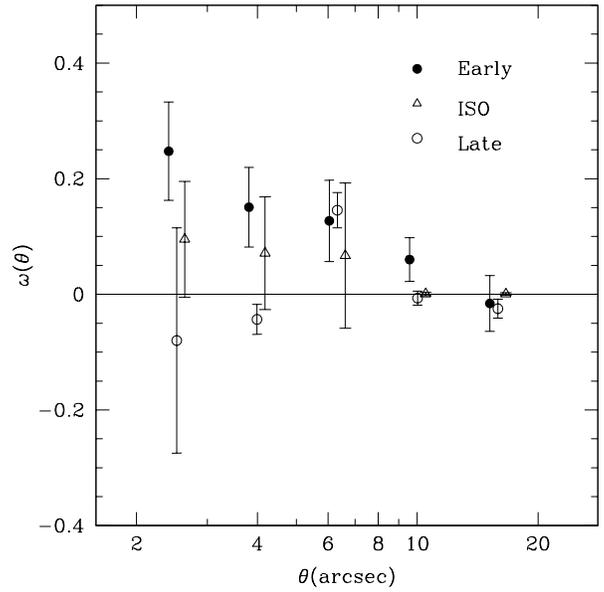}
\caption{Cross-correlation functions between particular galaxy targets and tracers galaxies with F606 $<$ 27}
\label{sfr}
\end{figure}

In order to compare these results with galaxy samples with well
determined characteristics, we have computed the angular
cross-correlation analysis between star-forming galaxies, early-type galaxies,
infrared galaxies
(ISO sources), and tracer galaxies in the Hubble Deep Field North.\\
For this purpose, we have considered
 blue spiral and irregular galaxies from \citet{rodipaper,rodicat}
with $H-K$ $<$ 0.7.
The sample of ellipticals was selected from \citet{stand03}.
The infrared selected galaxies was selected from ISOCAM observations in the Hubble Deep Field \citep{aussel1}. All sources were selected in the redshift range of 0.5 $<z<$ 1.2.
HDF tracer galaxies correspond to \citet{fer}, extracted in the F606W filter.
The results are plotted in Figure~\ref{sfr}, for tracer galaxies
with F606W $<$ 27.
We estimate correlation function error bars using uncertainties derived from
the Bootstrap re-sampling techniques.
We find a low cross-correlation amplitude at small angular scales, similar to
those obtained in our \hst/STIS images, in comparison to bright early-types galaxies and ISO sources in the same redshift range (see Figure~\ref{sfr}).
\\

\section{Discussion and conclusions}
We have analyzed different data sets corresponding to deep imaging in the
field of Gamma-Ray Burst hosts. A remarkable low correlation amplitude
at small angular scales is detected from the cross-correlation of  GRB's
and the surrounding galaxies in all samples analyzed.
The reliability of the results presented in this work can be judged
from the lack of correlation of stars in {\tt ISAAC} and \hst/STIS images.
By contrast, the galaxy-galaxy correlation function obeys the usual power-law
shape with a significant signal at small separations. Given that
these correlation functions were computed using angular positions,
any effect in real space would be diluted by projection of foreground
and background objects.

A comparison with a similar data set, centered in USS sources, indicates the very different 
environment of these two types of objects. While USS sources clearly reside in rich 
environments, GRBs are likely to reside in a typical or even lower galaxy density 
environment. Moreover the amplitude of the 
autocorrelation function of galaxies is larger than that of the GRB-galaxy cross-correlation. 

In the \hst/STIS fields 
GRB-tracers cross-correlation functions are significantly lower
than early-type-tracers cross correlations. In addition, GRB targets
have a lower cross correlation amplitude than ISO and early-type galaxy targets.
This suggests that the star formation events associated to GRBs occur in 
particularly low density environments, a result that
is supported by the fact that objects formed in global underdense
regions are expected to be biased to low luminosity, consistent
with GRB hosts characteristics \citep{floch03}.

\acknowledgements

This work was partially supported by the
Consejo Nacional de Investigaciones Cient\'{\i}ficas y T\'ecnicas,
Agencia de Promoci\'on de Ciencia y Tecnolog\'{\i}a,  Fundaci\'on Antorchas
and Secretaria de Ciencia y
T\'ecnica de la Universidad Nacional de C\'ordoba.
We acknowledge support from the SETCIP/CONICYT joint grant CH-PA/01-U01.
Dante Minniti is supported by FONDAP Center for Astrophysics 15010003.

\clearpage

\clearpage

\clearpage

\epsscale{0.5}

\clearpage
\begin{deluxetable}{ccccccc}
\tabletypesize{\footnotesize}
\tablecolumns{7}
\tablewidth{0pt}
\tablecaption{\label{tab1}GRB hosts  {\tt ISAAC} observations}
\tablehead{\colhead{Name} & \colhead{R.A. (J2000)} & \colhead{Dec. (J2000)} & \colhead{$z$} & \colhead{Reference}& \colhead{$K_{s}$$^1$} & \colhead{Limiting Magnitude} \\ & & & & & & \colhead{(1.5$\sigma$)} }
\startdata
GRB 981226 & $23^h$ $29^m$ $37^s.2$ & $-23\degr$ $55'$ $54"$ & $\sim1^\dagger$ &1 & 21.10$\pm$0.1& 24.4\\
GRB 990506 & $11^h$ $54^m$ $50^s.1$ & $-26\degr$ $40'$ $35"$ & 1.30  &2 & 21.45$\pm$0.2& 24.4 \\
GRB 990510 & $13^h$ $38^m$ $07^s.1$ & $-80\degr$ $29'$ $48"$ & 1.62  &3 & $\ge$22.5 & 22.5    \\
GRB 000210 & $01^h$ $59^m$ $15^s.5$ & $-40\degr$ $39'$ $33"$ & 0.85  &4 & 20.95$\pm$0.2 & 23.4\\
GRB 000418 & $12^h$ $25^m$ $19^s.3$ & $+20\degr$ $06'$ $11"$ & 1.12  &2 & 21.30$\pm$0.2 & 24.5\\
GRB 001011 & $18^h$ $23^m$ $04^s.6$ & $-50\degr$ $54'$ $16"$ &  $\sim1^\dagger$  & 1& 21.45$\pm$0.2& 23.4\\

\enddata

\tablerefs{
(1) \citealt{floch03}~;
(2) \citealt{Bloom02c}~;
(3) \citealt{Vreeswijk01}~;
(4) \citealt{Piro02a}~.}

\end{deluxetable}

\begin{deluxetable}{cccccccccc}
\tabletypesize{\tiny}
\tablecolumns{10}
\tablewidth{0in}
\tablecaption{\label{tab2}STIS {\tt 50CCD} GRB hosts sample characteristics}
\tablehead{\colhead{Name} & \colhead{Ref.} & \colhead{R.A. (J2000)} & \colhead{Dec. (J2000)} & \colhead{$z$} &  \colhead{Ref.} & \colhead{\tt CL} &\colhead{Ref.} & \colhead{exp. time} &\colhead{Limiting Mag.} \\ &&&&&&&&\colhead{(seconds)}& \colhead{(2$\sigma$)}}
\startdata
GRB 970228 & 1 & $05^h$ $01^m$ $46^s.7$& $+11\degr$ $46'$ $53"$ & 0.695 &20 &$25.8\pm0.25$&1 & 2300 &29.0\\
GRB 970508 & 2 & $06^h$ $53^m$ $49^s.5$& $+79\degr$ $16'$ $20"$ & 0.835 &21 &$25.25\pm0.20$&2 & 11688&30.3\\
GRB 971214 & 3 & $11^h$ $56^m$ $26^s.4$& $+65\degr$ $12'$ $01"$ & 3.42 &22 &$25.68\pm0.05$&(*)$^a$ &11874&29.7\\
GRB 980326 & 4 & $08^h$ $36^m$ $34^s.3$& $+79\degr$ $16'$ $20"$ & 1?	&23  &$29.25\pm0.25 $&4& 7200&27.5\\
GRB 980329 & 5 & $07^h$ $02^m$ $38^s.0$& $+38\degr$ $50'$ $44"$ &$<4$ &24 &28.6$\pm$0.3&5 & 8072&28.2\\
GRB 980519 & 6 & $23^h$ $22^m$ $21^s.5$& $+77\degr$ $15'$ $43"$ &$>1.5$ &25 &$27.0\pm0.2$& 37 & 8983&29.6\\
GRB 980613 &  7  &$10^h$ $17^m$ $57^s.6$ & $+71\degr$ $27'$ $26"$ & 1.096&26 &$26.3\pm0.1$&7 & 5851&28.7\\
GRB 980703 &  8 &$23^h$ $59^m$ $06^s.7$ & $+08\degr$ $35'$ $07"$ & 0.966&27 &$23.00\pm0.10$& 8 & 5178&29.2\\
GRB 981226 &  9 & $23^h$ $29^m$ $37^s.2$ & $-23\degr$ $55'$ $54"$& $\sim1^\dagger$ &28 &$25.04\pm0.07$&9  &8265&28.2\\
GRB 990123 &  10 &$15^h$ $25^m$ $30^s.3$ & $+44\degr$ $45'$ $59"$ & 1.6  &29 &$25.45\pm0.15$&10 & 7800&29.7\\
GRB 990308 &  11 &$12^h$ $23^m$ $11^s.4$ & $+06\degr$ $44'$ $05"$ & $>1.2?$&30 & $29.7\pm0.4$&37 &7842&29.0\\
GRB 990506 &  12 &$11^h$ $54^m$ $50^s.1$ & $-26\degr$ $40'$ $35"$ & 1.3	&31  &$24.5\pm0.1$ &(*)$^b$ &7914 &29.2\\
GRB 990510 &  13 &$13^h$ $38^m$ $07^s.3$ & $-80\degr$ $29'$ $49"$ & 1.619 &32   &$28\pm0.3$&13 &5840&29.2\\
GRB 990705 &  14 &$05^h$ $09^m$ $54^s.5$ & $-72\degr$ $07'$ $53"$ & 0.843 &33  &$22.45\pm0.10$& 33 &8851&29.6\\
GRB 990712 &  15 &$22^h$ $31^m$ $53^s.0$ & $-73\degr$ $24'$ $28"$ & 0.430 &32  &$\sim$23 &15 &4080& 29.5\\
GRB 991208 & 16 &$16^h$ $33^m$ $53^s.5$ & $+46\degr$ $27'$ $22"$ & 0.706&34 &$24.6\pm0.15$& 16  &5120 &30.0\\
GRB 991216 & 17& $05^h$ $09^m$ $31^s.3$ & $+11\degr$ $17'$ $07"$ & 1.02  &35 &$26.63\pm0.12$&(*)$^c$& 4744& 28.9\\
GRB 000418 & 18&$12^h$ $25^m$ $19^s.3$ & $+20\degr$ $06'$ $11"$ & 1.12 &31 &$24.30\pm$0.03&(*)$^a$ &5120& 29.5\\
GRB 000301 & 19&$16^h$ $20^m$ $18^s.6$ & $+29\degr$ $26'$ $36"$ & 2.04  &36 &$27.55\pm0.04$&(*)$^a$ &73911& 31.7\\

\enddata

\tablerefs{
(1) \citealt{fruch99a}~;
(2) \citealt{fruch151}~;
(3) \citealt{ode98}~;
(4) \citealt{fruch1029}~;
(5) \citealt{holland778}~;
(6) \citealt{holland698}~;
(7) \citealt{holland777}~;
(8) \citealt{holland00}~;
(9) \citealt{holland749}~;
(10) \citealt{Fruchter99a}~;
(11) \citealt{holland726}~;
(12) \citealt{hjorth731}~;
(13) \citealt{fruch757}~;
(14) \citealt{Holland00c}~;
(15) \citealt{fruch565}~;
(16) \citealt{fruch872}~;
(17) \citealt{fruch751}~;
(18) \citealt{fruch1061}~;
(19) \citealt{fruch1063}~;
(20) \citealt{bloom00}~;
(21) \citealt{Bloom98b}~;
(22) \citealt{kulkanat}~;
(23) \citealt{bloomnat}~;
(24) \citealt{lamb99}~;
(25) \citealt{jau01}~;
(26) \citealt{djo03}~;
(27) \citealt{djo98}~;
(28) \citealt{floch03}~;
(29) \citealt{hjorth219}~;
(30) \citealt{sha99}~;
(31) \citealt{Bloom02c}~;
(32) \citealt{Vreeswijk01}~;
(33) \citealt{floch02}~;
(34) \citealt{Castro-Tirado01}~;
(35) \citealt{Vreeswijk99a}~;
(36) \citealt{Jensen01}~;
(37) \citealt{jaunsen03}~;\\
(*) This Work. \\
$^{a, b, c}$  In a aperture of radius 1\farcs0, 2\farcs5, 0\farcs5.}

\tablecomments{ $\dagger$ : Derived from their $K$ magnitude and $R-K$ color.}
\end{deluxetable}

\end{document}